\documentclass[aps,prb,reprint,showpacs,groupedaddress]{revtex4-1}

\usepackage{bm}
\usepackage{wasysym}
\usepackage{amssymb}
\usepackage[cp1251]{inputenc}
\usepackage[dvips]{graphicx}
\usepackage[dvipsnames]{xcolor}
\begin{document}

\title{Interaction of gated and ungated plasmons in two-dimensional electron systems\\}

\author{A.A. Zabolotnykh}
\affiliation{Kotelnikov Institute of Radio-engineering and Electronics of the RAS, Mokhovaya 11-7, Moscow 125009, Russia\\
}
\affiliation{Moscow Institute of Physics and Technology, Institutskii per. 9, Dolgoprudny, Moscow region 141700, Russia\\}

\author{V.A. Volkov}
\email{Volkov.V.A@gmail.com}
\affiliation{Kotelnikov Institute of Radio-engineering and Electronics of the RAS, Mokhovaya 11-7, Moscow 125009, Russia\\
}
\affiliation{Moscow Institute of Physics and Technology, Institutskii per. 9, Dolgoprudny, Moscow region 141700, Russia\\}

\date{\today}

\begin{abstract}
	Unique properties of plasmons in two-dimensional electron systems (2DESs) have been studied for many years. Existing theoretical approaches allow for analytical study of the properties of ungated and gated plasmons in two fundamental, ideal cases - the 2DES in dielectric environment and under an infinite metallic gate, respectively. However, it is for the first time that we introduce an analytical theory of the interaction of gated and ungated plasmons in partly gated 2DES. Generally, a finite-width gate is formed by a metallic strip placed over an infinite plane hosting 2D electrons. Our solution, in particular, describes the propagating plasmon modes with their charge density having $N$ nodes under the gate. In this regard, a new mode with $N=0$ has been found in addition to the gapped modes with $N =1, 2,...$ previously derived from numerical calculations. Unexpectedly, this fundamental plasmon mode has been found to differ substantially from the rest. In fact, it is characterised by gapless square root dispersion and represents a hybrid of gated and ungated plasmons. In contrast to the higher modes, the currents and lateral fields of the fundamental mode are localized mainly to the outside area in the vicinity of the gate. Heretofore, such a 'near-gate plasmon' has never been considered.
\end{abstract}

\maketitle

\section{Introduction}
Plasma oscillations (plasmons) in 2DES  are radically different from conventional 3D plasmons. When 2DES is embedded in a dielectric medium with permittivity $\varkappa$, the 2D plasmons can be characterized by the following square root dispersion law\cite{Stern}: 
\begin{equation}
 \label{plasmon}	
 \omega_{p}(q)=\sqrt{\frac{2\pi n e^2q}{\varkappa m}}, \quad q=\sqrt{q_x^2+q_y^2},
\end{equation} 
where $n$ is the 2D electron concentration,  
$m$ is the electron effective mass, and $q$ is the 2D wave vector of the plasmon. 

Furthermore, screening electron-electron ($e$-$e$) interaction with a metallic gate leads to reduction in the ungated plasmon frequency (\ref{plasmon}) by the factor of $\sqrt{2dq}$, where $d$ is the distance between the gate and 2DES. Consequently, at long wavelengths ($qd\ll 1$) the dispersion for the gated plasmons becomes linear\cite{Chaplik}:
\begin{equation}
\label{screened}
	\omega_g(q)=qV_p,\quad V_p= \sqrt{\frac{4\pi n e^2d}{m\varkappa}},
\end{equation}
where $V_p$ is the velocity of gated plasmons.

Originally 2D plasmons were observed in 2D systems of electrons on a liquid helium surface\cite{Grimes} and in silicon inversion layers\cite{Allen,Theis_1977}. Thus far, plasmons have been studied in different 2D electron systems including semiconductor heterojunctions and quantum wells\cite{Lusakowski_2017,Muravev_2011,Scalari_2012,Muravev_2015_ani,
Muravev_2016,Grigelionis_2015,Gusikhin_Muravev,Andreev_2017,Muravev_2017}, graphene\cite{Woessner_2014,Chen_2012,Fei_2012,Crassee_2012}, topological isolators\cite{Politano_2015,Song_2016,Kumar_2016}, transition metal dichalcogenide monolayers\cite{Enaldiev_2018}, {\it etc}.

Dependence of plasmon frequency on 2D electron concentration, $n$, permits easy tuning of the 2D plasmons over a wide frequency range by means of a gate voltage control. For this reason, 2D gated structures are proving promising not only in fundamental studies of collective excitation physics, but also as detectors and emitters of electromagnetic radiation in the terahertz range\cite{Dyakonov_1993,Dyakonov_2010,Dyer_2012,Aizin_2006,Popov_Fateev,Aizin_2007,
Sydoruk_2015,Muravev,Tsui_1980,Kock_1988,Steinmuller_1994,Satou_2005,Satou_Khmyrova_2005,
Satou_2006}.

When a perpendicular magnetic field is imposed on a restricted 2DES (with or without a gate), the two types of plasmons emerge - the bulk magnetoplasmons with a spectrum gap and the gappless, one-way, edge magnetoplasmons (EMPs). Both the exact solution\cite{Volkov_1985,Volkov_1988} and the modeling approximation\cite{Fetter_1985} show that in the case of a semi-planar form of the 2DES, EMPs can exist at any magnetic field strength regardless of the metallic gate. What is more, in a classically strong magnetic field, EMPs are subject to insignificant damping, even in low-mobility samples. It is these properties that have been stimulating the EMP research for a number of years\cite{Kukushkin_2004,Muravev_2015,Gusikhin_2015,Gusikhin_2018,Principi_2016,Bosco_2017}. In addition, it has recently been shown\cite{Jin_2016} that the gated magnetoplasmons may belong to one of the 2D classes within the framework of topological classification of bosons\cite{Kitaev_2009}, and that EMPs are topologically protected. These findings have generated a great deal of interest in 2D plasmon physics as well.

Due to their peculiar nature, mathematical analysis of 2D plasmons has always been rather challenging. Considering a confined 2DES, for example, in the form of a half-plane, a strip or a disk, the relationship between the charge density and the induced electric field is nonlocal. Therefore, the calculation of the confined plasmon spectrum involves finding a solution to a complicated integral equation, which, in fact, can be solved analytically only for several oversimplified cases\cite{Fetter_1986,Leavitt_1986,Nazin_1987,Cataudella_1987,Giovanazzi_1994,
Aleiner_1994,Aleiner_1995,Mishchenko_2010}.

\begin{figure}
			\includegraphics[width=7.0cm]{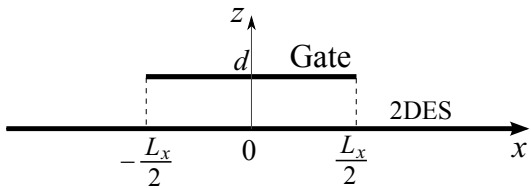}
			\caption{ \label{Fig:Geom} System under consideration with the gate and the 2DES assumed to be infinite in $y$.
}
\end{figure}

In this paper we report the analytical investigation of the influence the finite strip-shaped metallic gate has on the 2D plasmon spectrum in infinite 2DES. In addition to being interesting in itself from the fundamental standpoint, the given geometry, shown in Fig.~\ref{Fig:Geom}, represents an elementary constituent of a metallic grating that can be applied to excite 2D plasmons\cite{Allen}. Similar configurations have been implemented in detectors and emitters of electromagnetic radiation. 

As a rule, the 2DES with a finite gate is examined by means of numerical methods\cite{Iranzo_2018,Popov_2008,Satou_2003,Satou_2004,Ryzhii_2006,Petrov_2017,Popov_2005,
Davoyan_2012,Bylinkin_2018}, which generally includes calculation of the absorption of electromagnetic wave with its electric field component directed across a long metal strip. In this case, however, the plasmon modes governed by gapless dispersion law, even if they exist, are not expressed. In the course of our analysis, we prove that it is the fundamental mode that has this feature.

In the system under consideration, it has been assumed, as a matter of convenience, that $e$-$e$ interaction of the gate electrons is screened by the electrons of 2DES, in contrast to the case of a finite 2DES with an infinite gate, where $e$-$e$ interaction of electrons in 2DES is screened by the gate electrons instead. In this particular instance, it is also convenient to solve the integral equation for the charge electron density in the gate. The analytical solution to the problem is found based on two realistic assumptions: 1) the sought-for frequency of the new plasmon mode is small compared to that of the ungated plasmon, and 2) the distance between the gate and 2DES $d$ is small compared to the gate size $L_x$ and the plasmon wavelength ($d\ll L_x,q^{-1}$). Given these approximations and an assumption of sufficiently large gate conductivity, see Sec.~\ref{SDis}, the exact integral equation for the plasmon charge density can be reduced to a differential equation with boundary conditions specified at the edges of the gate. The resultant solutions describe the plasmons which propagate along the gate and are confined to the regions under and near the gate. The spectrum of these plasmons consists of a series of 1D sub-bands $\omega_N(q_y)$, shown in Fig.~\ref{Fig:Spectrum}, where $N$ denotes an integer number of half-wavelengths across the strip, $N=0,1,2,...$

For the fundamental mode ($N=0$) with  $|q_y|L_x\ll 1$, the plasmon spectrum is defined by:
\begin{equation}
\label{disp_fund}
	\omega_0(q_y)=\sqrt{\frac{8\pi e^2 n d}{m\varkappa}\frac{|q_y|}{L_x}}.
\end{equation}
The dispersion of the fundamental mode is rather unusual as it combines the features of both ungated (\ref{plasmon}) and gated (\ref{screened}) plasmons. The currents in this mode are localized mainly to the outside region near the gate. We shall further refer to the mode in (\ref{disp_fund}) as to the ''near-gate'' plasmon. Eq.~(\ref{disp_fund}) is one of the key results of this paper. The influence of the external perpendicular magnetic field on the fundamental mode $\omega_0(q_y)$ is analyzed in Sec.~\ref{SDis}.

\section{Plasmons in 2DES with a strip-shaped gate}  
The following discussion pertains to the system geometry in Fig.~\ref{Fig:Geom}, with an infinite 2DES in $z=0$ plane and a metallic gate at the distance $d$ above it. The gate is infinitely long in $y$, has a finite width defined by $\left[ -L_x/2,L_x/2 \right]$ in $x$ and conductivity $\sigma_g$.

The desired solutions are restricted to the waves propagating along the gate according to  $\exp(iq_y y-i\omega t)$. We consider the spectra in the long-wavelength limit $|q_y|\ll k_F$, where $\hbar k_F$ is the Fermi momentum and take the classic approach (Ohm's law with the collisionless Drude model for conductivity of 2DES) to describe the electron dynamics. We also neglect the electromagnetic retardation effects.

The Poisson equation for the plasmon potential $\varphi(x,z)$ can be formulated as follows:
\begin{equation}
\label{Poisson}
	(\partial_x^2+\partial_z^2-q_y^2)\varphi(x,z)= -\frac{4\pi}{\varkappa}\left[\rho(x)\delta(z)+\rho_g(x)\delta(z-d)\right],
\end{equation}
where we assume the 2DES and the metallic gate have infinitesimal thickness; $\rho(x)$ and $\rho_g(x)$ are the plasmon charge densities in 2DES and in the gate, respectively; $\rho_g(x)$ equals zero outside the strip.
 
Using Green's function Eq.~(\ref{Poisson}) can be put in the following equivalent form:
\begin{eqnarray}
 \label{Green}
 	\varphi(x,z)= \int^{\infty}_{-\infty} G(x-x',z)\rho(x')dx'+ \nonumber & \\
 	\int^{\infty}_{-\infty} G(x-x',z-d) \rho_g(x')dx',&
\end{eqnarray}
where $G(x,z)=2K_0(|q_y|\sqrt{x^2+z^2})/\varkappa$, $K_0(x)$ is the modified Bessel function of the second kind and zeroth order.
 
If we define the plasmon potential in 2DES and in the gate as $\varphi(x)=\varphi(x,0)$ and  $\varphi_g(x)=\varphi(x,d)$, respectively, the Fourier transformation of Eq.~(\ref{Green}) will lead to:
 \begin{equation}
 \label{system_pot}
 	\begin{array}{lcr}
    \varphi(q_x)=\frac{2\pi}{\varkappa\sqrt{q_y^2+q_x^2}}\left(\rho(q_x)+\rho_g(q_x)e^{-d\sqrt{q_y^2+q_x^2}}\right),\\
   \varphi_g(q_x)=\frac{2\pi}{\varkappa\sqrt{q_y^2+q_x^2}}\left(\rho(q_x)e^{-d\sqrt{q_y^2+q_x^2}}+\rho_g(q_x)\right).
  \end{array}
\end{equation}
 
The Ohm's law and the continuity equation can be used to derive the relation between $\varphi(q_x)$ and  $\rho(q_x)$ as follows:
\begin{equation}
\label{cont_q}
	i\omega\rho(q_x)=\sigma(q_x^2+q_y^2)\varphi(q_x),
\end{equation}
 where $\sigma=\sigma(\omega)$ is the dynamic conductivity of 2DES. 

Eliminating $\varphi(q_x)$ and $\rho(q_x)$ in Eqs.~(\ref{system_pot}) and (\ref{cont_q}), and then taking the inverse Fourier transform leads to:
\begin{eqnarray}
\label{int_eq}
	\varphi_g(x)=\frac{\alpha}{\varkappa}\int^{+\infty}_{-\infty} \frac{e^{i q_x x} e^{-2 d\sqrt{ q_y^2+ q_x^2}}\rho_g( q_x)} {1-\alpha\sqrt{ q_y^2+ q_x^2}}d q_x + \nonumber & \\ 
	\frac{2}{\varkappa}\int^{L_x/2}_{-L_x/2} K_0( |q_y| | x- x'|)\rho_g( x')d x', &
\end{eqnarray}
where  $\sigma=e^2n/(-i\omega m)$, based on collisionless Drude model for conductivity, and $\alpha=2\pi e^2 n/(\varkappa m\omega^2)$.

Since the plasmons of interest are coupled to the gate and their spectrum lies ''outside'' the spectrum of the bulk plasmons existing far from the gate, their frequency, $\omega$, is lower than that of the bulk plasmon, $\omega_p(q_y)$, for the same wave vector, $q_y$. This condition can be stated as $\alpha |q_y|> 1$. Thus, the denominator in the first integral in Eq.~(\ref{int_eq}) does not go to zero.

For the case of $\alpha |q_y|\gg 1$, we can introduce the following expansion:
\begin{equation}
\label{expansion}
	\frac{\alpha}{1-\alpha \sqrt{ q_y^2+ q_x^2}}=\sum_{M=0}^{\infty}\frac{-1}{\alpha^M( q_y^2+ q_x^2)^{(M+1)/2}}.
\end{equation}
In the series above, it is sufficient to keep only the first two dominant terms with $M=0,1$. Hence, after this approximation, Eq.~(\ref{int_eq}) takes the form of:
\begin{eqnarray}
\label{simple_int}
	\varphi_g(x)+\frac{1}{\varkappa}\int^{+\infty}_{-\infty} e^{i q_x  x} \rho_g( q_x) \frac{e^{-2 d\sqrt{ q_y^2+ q_x^2}}}{\alpha( q_y^2+ q_x^2)}d q_x =& \nonumber\\
	\frac{2}{\varkappa}\int^{L_x/2}_{-L_x/2} 	\Delta K( x - x')\rho_g( x')d x',&
\end{eqnarray}
where
\begin{equation}
	\Delta K(x)=K_0( |q_y| | x|)-K_0( |q_y| \sqrt{ x^2+4d^2}).
\end{equation}
Next, we can make further approximations of $d\ll L_x$ and $|q_y|d\ll 1$. In these limits, $\Delta K(x)$ becomes a $\delta$-function\cite{Chaplik}, $C\delta(x)$, with coefficient $C$  defined by the integrated area of $\Delta K(x)$. In fact, it can be found that $C=2\pi d$, which is the so-called local capacity approximation. Moreover, on the left-hand side of Eq.~(\ref{simple_int}) the factor $\exp(-2 dq)$ can be reduced to one.

 Based on these additional assumptions, Eq.~(\ref{simple_int}) can be modified as:
\begin{equation}
\label{eq_int_q}
	\varphi_g(x)+\frac{1}{\varkappa\alpha}\int^{+\infty}_{-\infty} e^{i  q_x  x}  \frac{\rho_g( q_x)}{ q_y^2+ q_x^2}d q_x =\frac{4\pi d}{\varkappa}\rho_g( x),
\end{equation}
where $-L_x/2 \le x \le L_x/2$.

Finally, expressing $\rho_g(q_x)$ in terms of $\rho_g(x)$ allows Eq.~(\ref{eq_int_q}) to be rewritten in a more convenient way as:
\begin{equation}
\label{fin_int}
	\frac{4\pi d}{\varkappa} \rho_g(x)=\frac{\pi}{|q_y|\varkappa\alpha}\int^{L_x/2}_{-L_x/2} e^{- |q_y| | x- x'|} \rho_g( x')d x'+\varphi_g(x).
\end{equation}

Here, it is worth noticing that since the gate conductivity, $\sigma_g$, is typically large enough, the plasmon potential,  $\varphi_g(x)$, inside the metal strip becomes negligible compared to the other two terms in Eq.~(\ref{fin_int}). More detailed discussion on estimation of $\sigma_g$ and $\varphi_g(x)$ is presented in Sec.~\ref{SDis}. 

In fact, the assumption of negligible $\varphi_g(x)$ is crucial to finding the exact solution to (13) as it enables the reduction of the integral equation to the differential form\cite{Manzhirov}:

\begin{equation}
\label{dif_eq}
	\left(\partial_x^2-q_y^2+\frac{\omega^2}{V_p^2}\right)\rho_g(x)=0
\end{equation}
with the boundary conditions specified as: 
\begin{equation}
\label{bc}
 	\begin{array}{lcr}
    \Big(\partial_x\rho_g(x)-|q_y|\rho_g(x)\Big)|_{x=-L_x/2}=0,\\
   \Big(\partial_x\rho_g(x)+|q_y|\rho_g(x)\Big)|_{x=L_x/2}=0.
  \end{array}
\end{equation}
 
At this stage, we can also introduce the effective transverse wave number, $k$, to replace the non-conserved transverse wave vector, $q_x$, according to:
\begin{equation}
	k^2=\frac{\omega^2}{V_p^2}-q_y^2,
\end{equation}
where $k$ takes a discrete value for a given $q_y$, as shown below.

The solutions to Eqs.~(\ref{dif_eq}) and (\ref{bc}) have a certain parity since the even and odd solutions have the form of $\cos k x$ and $\sin k x$ with $k$ defined by the dispersion relation as follows:
\begin{equation}
\label{odd_even}
	k\left(\tan \frac{kL_x}{2}\right)^{\pm 1}=\pm |q_y|,
\end{equation}
where the  $'+'$ and $'-'$ signs correspond to the even and odd modes, respectively.

It can be noted that the dispersion equation for odd modes has a trivial solution for $k=0$, which can be disregarded as it implies zero charge density, $\rho_g(x)=0$.

\begin{figure}
			\includegraphics[width=8.0cm]{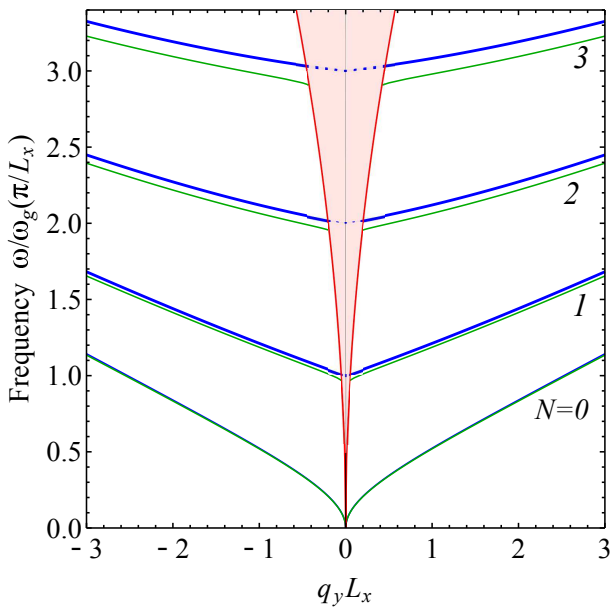}
			\caption{ \label{Fig:Spectrum} Spectrum of the 2D plasmons in 2DES with the gate in the form of a strip, obtained analytically (in blue) and numerically (in green). The fundamental mode $N=0$ is a hybrid near-gate plasmon. Also shown  (in red) is the boundary of the bulk plasmon continuum, $\omega=\omega_p(q_y)$, within which the excited higher modes $N=1,2,...$ have a finite lifetime. The parameter $\sqrt{d/L_x}$ was assumed to have the value 0.05.
}
\end{figure}

From Eq.~(\ref{odd_even}) we obtain a discrete series of the plasmon modes  $N=0,1,2,...$ with mode frequencies, $\omega_N(q_y)$, and transverse wave numbers, $k_N(q_y)$, as illustrated in Figs.~2 and 3. For $|q_y|L_x\ll 1$ the fundamental $N=0$ mode has an unusual square root dispersion (\ref{disp_fund}), in contrast to the linear dispersion in 2DES with an infinite gate\cite{Chaplik}. Note that this mode has frequency much lower than $\omega_p(q_y)$, therefore condition $\alpha |q_y| \gg 1$ is satisfied.

The plasmon spectrum outside the bulk continuum was also determined numerically. By first expanding $\rho_g(x)$ in Eq. (\ref{int_eq}) into its series form, with $\sin(\pi P x/L_x)$, $P=1,3,5,..$ and $\cos(\pi P x/L_x)$, $P=0,2,4,..$ denoting the odd and the even modes, correspondingly and then following a standard computational procedure, we arrived at the spectrum plotted in green in Fig.~\ref{Fig:Spectrum}. It is clear that for the fundamental mode the numerical and analytical solutions match perfectly. For the higher excited modes $N=1,2,3,..$ the results show close agreement overall, though numerical solution yields slightly lower frequencies.

The analytically obtained modes with $N>0$ have nonzero frequencies at $q_y=0$, thus they lie inside the bulk spectrum (\ref{plasmon}) and strongly interact with the continuum of ungated plasmons. Analogous, for instance, to the inter-edge magnetoplasmons\cite{Mikhailov_1992}, such an interaction can lead the decay of the gated modes and to appearance of small imaginary corrections to the plasmon frequency, due to the finite lifetime of exited modes.

If one neglects these corrections, then these modes have long wavelength asymptotes:
\begin{equation}
	\omega_N^2/V_p^2=[\pi^2 N^2/L_x^2]+b_N|q_y|/L_x,
\end{equation} 
where $|q_y|L_x\ll 1$, $b_N=2$ for $N=0$ and $b_N=4$ for $N>0$;
i.e. at $q_y=0$ the frequency $\omega_N$ equals the frequency of the gated plasmon (\ref{screened}) according to the common ''quantization rule'' $q_x\rightarrow \pi N/L_x$.

For $|q_y| L_x \gg 1$, all the modes have asymptotic behavior described by: $\omega_N^2/V_p^2=[\pi^2(N+1)^2/L_x^2]+q_y^2$, i.e. $\omega_N$ tends to the frequency of the gated plasmon (\ref{screened}) as $q_x \rightarrow \pi(N+1)/L_x$.

\begin{figure}
			\includegraphics[width=7.0cm]{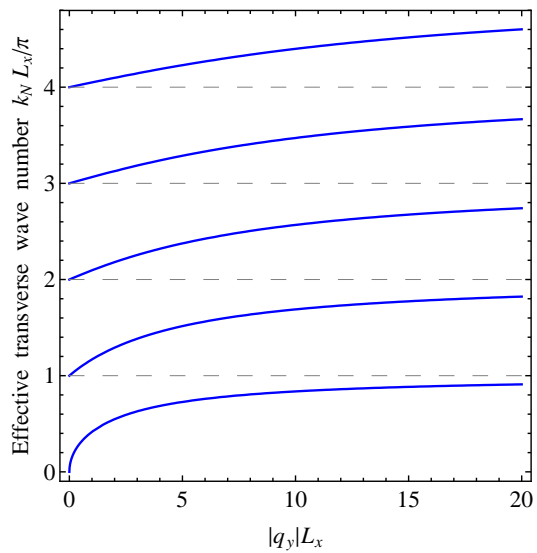}
			\caption{ \label{Fig:Kx} Quantization rule for the transverse wave number $k_N$ as a function of the wave vector along the strip $q_y$. The curves are universal and do not depend on parameters of the system.
}
\end{figure}

The predicted spectrum $\omega_N(q_y)$ can be understood in terms of relation (\ref{screened}), provided that $q_x$ is replaced by effective transverse wave number $k_N$, defined by Eqs.~(\ref{odd_even}). Fig.~\ref{Fig:Kx} illustrates the dependence of $k_N$ on $q_y$, where the curves are plotted in dimensionless variables $(kL_x,q_yL_x)$ and do not depend on parameters of the system. 

\begin{figure}
			\includegraphics[width=8.0cm]{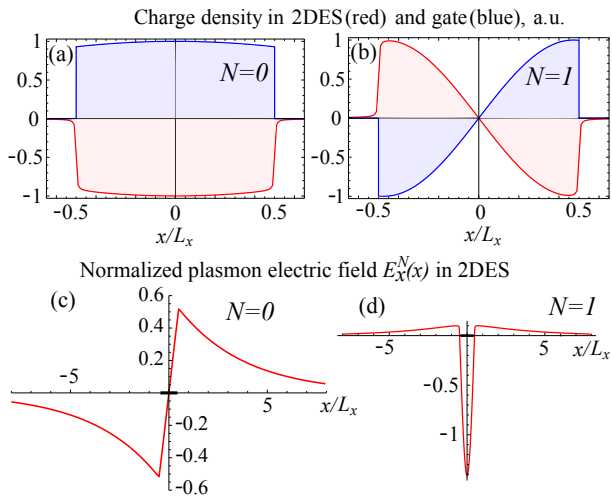}
			\caption{ \label{Fig:Field_all} Plasmon charge density (a,b) and normalized electric field $E_x^N$ (c,d) in 2DES for the modes $N=0$ and $N=1$  as a function of $x/L_x$. In (a) and (b), the charge densities in 2DES and in the gate are denoted by red and blue traces. In (c) and (d) the electric field is normalized by $\sqrt{\int^{+\infty}_{-\infty}E_x^2(x) dx/L_x}$. Other parameters were set as follows: $q_y L_x=0.3$, $\sqrt{d/L_x}=0.05$. The gate is bounded in $x$ direction by the interval $[-1/2, 1/2]$.
}
\end{figure}

Figure~\ref{Fig:Field_all} displays the plasmon charge density distribution in 2DES and in the gate, and the electric field component $E_x$ in 2DES for the modes $N=0,1$. From these data it is evident that the plasmon charge density in 2DES is localized entirely to the region under the gate. For $|x|\gg L_x$ and $|q_y| L_x  \ll 1$, the electric field, $E_x(x)$, in 2DES decreases with the characteristic length of the order of $|q_y|^{-1}$. For the parameters indicated in Figs.~\ref{Fig:Field_all}\,(c) and 4(d), the typical ratio of normalized electric fields $E_x^N$ under the gate for $N=0$ and $N=1$ is approximately $1/3$ and tends to zero as $|q_y|L_x \to 0$.

\section{Discussion}\label{SDis}

In the course of our investigation we obtained analytical spectra for the condition of $\alpha |q_y|\gg 1$, as stated in (9). However, this assumption becomes invalid for the modes $N=1,2,...$ both inside and in the vicinity of the bulk continuum. Nevertheless, as demonstrated in Fig.~\ref{Fig:Spectrum}, the analytical and numerical solutions have been found to be in close agreement, even in the vicinity of the bulk continuum. In this regard, first, it should be noted that in fact, the expansion in (\ref{expansion}) is valid when $\alpha|q|\gg 1$, where $|q|=\sqrt{{\tilde q_x}^2+q_y^2}$ and $\tilde q_x$ takes the typical values of $q_x$ with the greatest contribution to the first integral in Eq. (\ref{int_eq}). It is clear from Fig.~\ref{Fig:Kx} that $\tilde q_x$ is of the order of $k_N(q_y)$. Consequently, we find that for the estimated dispersion curves (plotted in blue in Fig.~\ref{Fig:Spectrum}) the condition $\alpha\sqrt{q_y^2+k_N^2(q_y)}\gg 1$ is satisfied, even for the data representing the region inside the bulk continuum. Therefore, the obtained analytical solution remains valid (accurate to small corrections, that may take imaginary values), even when the condition $\alpha |q_y|\gg 1$ is violated.

Next, we identify the conditions under which the plasmon potential in the gate, $\varphi_g(x)$, in Eq.~(\ref{fin_int}) can be neglected in comparison with $4\pi d \rho_g(x)/\varkappa$. In this case, by applying the continuity equation within the gate, the $\varphi_g$ and the corresponding gate conductivity, $\sigma_g$, can be estimated as:
\begin{equation}
\varphi_g\sim\left|\frac{i\omega\rho_gL_x^2}{\sigma_g(1+q_y^2L_x^2)}\right|, \quad 
\sigma_g\gg\frac{\omega L_x^2\varkappa}{4\pi d(1+q_y^2L_x^2)}.
\end{equation}
Moreover, it can be noted that in the limit $|q_y|L_x\ll 1$, the result in (19) has a straightforward physical interpretation. If we define the capacitance $\widetilde C=L_x L_y\varkappa/(4\pi d)$, the resistance $R=L_x/(\sigma_g L_y)$ and $L_y\approx \pi/q_y$, then, according to (19), the time it takes the capacitor to discharge through $R$ should be much less than the plasma oscillation period $2\pi/\omega$. 

In the given analysis we did not take into account the electromagnetic retardation effects, thus, the produced results are applicable only for $\omega\ll cq/\sqrt{\varkappa}$, where $c$ is the speed of light in vacuum.

Thus far, in our analysis we have made no allowance for any external magnetic field applied to the system. Therefore, assuming that gate conductivity, $\sigma_g$, does not depend on magnetic field, the special case of interest to consider is when the system is placed in the perpendicular magnetic field $B$. Provided such a condition, the conductivity, $\sigma$, in Eq.~(\ref{cont_q}) is  replaced with  the diagonal conductivity, $\sigma_{xx}(\omega,B)$. Consequently, in the collisionless approximation, parameter $\alpha$ in Eq.~(\ref{int_eq}) takes the following form:
\begin{equation}
	\alpha=\frac{2\pi e^2 n}{\varkappa m(\omega^2-\omega_c^2 )}
\end{equation}
where $\omega_c=|e|B/(mc)$ is the electron cyclotron frequency.
Hence, $\omega^2$ in (8) is now replaced with $\omega^2-\omega_c^2$, yielding the \textit{B}-dependent plasmon spectrum $\omega_N(B)=\sqrt{\omega_N^2+\omega_c^2}$.

Regarding the magnetic field effect discussion, it is worth mentioning that we did not find EMP modes\cite{Fetter_1985,Volkov_1985} as their existence is contingent on the condition of inhomogeneous Hall conductivity, which is not the case in our system.

Thus, we show that the metallic gate itself confines plasmons, even without any changes in electron density within 2DES, i.e., when 2DES is homogeneous. This conclusion is in qualitative agreement with the results in Refs.\cite{Iranzo_2018,Bylinkin_2018} obtained numerically for several excited modes in graphene with metallic grating and other 2DESs\cite{Popov_2005,Davoyan_2012}.  

Finally, as a matter of practical example, we estimate the fundamental mode numerics for the following characteristic parameters of the 2DES formed by GaAs/AlGaAs quantum well: the electron concentration $n=3\cdot 10^{11}$ cm$^{-2}$, the average dielectric permittivity $\varkappa = 7$, the mobility $\mu$ = $10^5$ cm$^2/$(V\,s) at 77 K and $10^4$ cm$^2/$(V\,s) at 300 K. For the gate dimensions, $d=200$ nm, $L_x=1$ $\mu$m and $L_y=10$ $\mu$m we find $\omega/(2\pi)=0.25$ THz and the following quality factors: $\omega\tau=6$ at 77 K and $\omega\tau=0.6$ at 300 K. If we take into account the fact that the electromagnetic retardation can significantly increase the quality factor of plasmons\cite{Gusikhin_2018,Kukushkin_2006,Kukushkin_2003}, then plasmons can be well-defined even at room temperatures.

\section{Conclusion}

In summary, we have examined analytically the plasmon modes in 2DES with the gate formed by a metallic strip over a 2D electron plane. The oscillating charge of these modes is found to be confined under the gate. The plasmon spectrum has been characterized by the mode number $N = 0, 1, 2,...$ and the wave vector along the gate $q_y$. Higher modes ($N=1,2,...$) are known from numerical calculations and possess the gapped dispersion law. In addition, a new mode ($N=0$) is found. This fundamental mode is a hybrid of gated ($\omega_0\propto|q_y| \sqrt{d}$) and ungated ($\omega_0\propto \sqrt{|q_y|}$)  plasmons. It did not appear in numerical calculations of the THz absorption due to its gapless spectrum. Its currents and fields are localized mainly to the outside region near the gate. Up until now, this "near-gate plasmon" has never been considered.
The obtained spectra can be interpreted in terms of the gated plasmon spectrum (\ref{screened}), in which $q_x$ is  replaced by the quantized wave number $k_N$, see Fig.~\ref{Fig:Kx}, $k_N$ lies between $\pi N/L_x$ and $\pi (N+1)/L_x$. Our findings are promising for possible applications in integral sub-THz optics and nanoplasmonics, for example, to transmit sub-THz signals within the integrated circuits.

\begin{acknowledgments}
We thank I. V. Kukushkin and V. M. Muravev for stimulating discussion of unpublished experimental data. Numerous constructive discussions of one of the authors (V.V.) with I. Kukushkin have resulted in the formulation of the considered problem. The work was done within the framework of the state order and supported by the Russian Foundation for Basic Research (project No. 17-02-01226).
\end{acknowledgments}

\end{document}